\documentclass[conference]{IEEEtran}

\usepackage{acronym}
\usepackage{amsfonts}
\usepackage[dvips]{graphicx}
\usepackage{times}
\usepackage{cite}
\usepackage{amsmath}
\usepackage{array}
\usepackage{amssymb}
\usepackage{stfloats}
\usepackage{diagbox}
\usepackage{graphicx}
\usepackage{footnote}
\usepackage{amsthm}
\usepackage{booktabs}
\usepackage{array}
\usepackage[ruled,vlined]{algorithm2e}
\usepackage{subeqnarray}
\usepackage{cases}
\usepackage{threeparttable}
\usepackage{color}
\usepackage{epstopdf}
\usepackage{makecell}
\usepackage{multirow}
\usepackage{tabularx}
\usepackage{enumerate}
\usepackage{multicol}
\usepackage{subfig}
\usepackage{caption}

\def\bb0{{\mathbb{0}}}

\def\bb{{\mathbf{b}}}

\def\bm{{\mathbf{m}}}

\def\b0{{\mathbf{0}}}

\def\bbC{{\mathbb{C}}}

\def\sf0{{\mathsf{0}}}

\def\rm0{{\mathrm{0}}}

\acrodef{CSI}[CSI]{channel state information}
\acrodef{CSIT}[CSIT]{channel state information at the transmitter}
\acrodef{CSIR}[CSIR]{channel state information at the receiver}
\acrodef{MIMO}[MIMO]{multiple-input multiple-output}
\acrodef{SISO}[SISO]{single-input single-output}
\acrodef{MISO}[MISO]{multiple-input single-output}
\acrodef{SIMO}[SIMO]{single-input multiple-output}
\acrodef{ADCs}[ADCs]{analog-to-digital convertors}
\acrodef{SNR}[SNR]{signal-to-noise ratio}
\acrodef{AWGN}[AWGN]{additive white Gaussian noise}
\acrodef{MRT}[MRT]{maximal ratio transmission}
\acrodef{DFT}[DFT]{Discrete Fourier Transform}
\acrodef{ULA}[ULA]{uniform linear array}
\acrodef{UPA}[UPA]{uniform planar array}
\acrodef{LS}[LS]{least squares}
\acrodef{ALMMSE}[ALMMSE]{approximate linear minimum mean squared error}
\acrodef{QIHT}[QIHT]{quantized iterative hard thresholding}
\acrodef{QIST}[QIST]{quantized iterative soft thresholding}
\acrodef{SVD}[SVD]{singular value decomposition}

\usepackage{hyperref}
\usepackage{bm}
\usepackage{threeparttable}
\hypersetup{draft}

\ifCLASSINFOpdf

\else
\fi
\begin{document}

\title{Adaptive Channel Estimation Based on Model-Driven Deep Learning for Wideband mmWave Systems
\thanks{This work was supported in part by the National Natural Science Foundation of China under Grant 61941104. }
}

\author{
\IEEEauthorblockN{Weijie Jin\IEEEauthorrefmark{1}, Hengtao He\IEEEauthorrefmark{1},  Chao-Kai Wen\IEEEauthorrefmark{2}, Shi Jin\IEEEauthorrefmark{1}, Geoffrey~Ye~Li\IEEEauthorrefmark{3}}
\IEEEauthorblockA{\IEEEauthorrefmark{1}National Mobile Communications Research Laboratory, Southeast University\\ Nanjing 210096, P. R. China, E-mail: \{jinweijie, hehengtao, jinshi\}@seu.edu.cn}
\IEEEauthorblockA{\IEEEauthorrefmark{2}Institute of Communications Engineering, National Sun Yat-sen University\\ Kaohsiung 804, Taiwan, E-mail: chaokai.wen@mail.nsysu.edu.tw}
\IEEEauthorblockA{\IEEEauthorrefmark{3}Department of Electrical and Electronic Engineering, Imperial Colledge London\\ London, UK, E-mail:
geoffrey.li@imperial.ac.uk}
}

\maketitle

\begin{abstract}
Channel estimation in wideband millimeter-wave (mmWave) systems is very challenging due to the beam squint effect. To solve the problem, we propose a learnable iterative shrinkage thresholding algorithm-based channel estimator (LISTA-CE) based on deep learning. The proposed channel estimator can learn to transform the beam-frequency mmWave channel into the domain with sparse features through training data. The transform domain enables us to adopt a simple denoiser with few trainable parameters. We further enhance the adaptivity of the estimator by introducing \emph{hypernetwork} to automatically generate learnable parameters for LISTA-CE online. Simulation results show that the proposed approach can significantly outperform the state-of-the-art deep learning-based algorithms with lower complexity and fewer parameters and adapt to new scenarios rapidly.
\end{abstract}


\IEEEpeerreviewmaketitle

\section{Introduction}
\par
Beamspace multiple-input multiple-output (MIMO) implemented by a discrete lens antenna array is a promising technique for 5G wireless communications. It transforms the spatial channel into the beam domain to concentrate the signal from different directions\cite{Zeng2014}, which can reduce the number of radio-frequency (RF) chains significantly and thus has been successfully employed in millimeter-wave (mmWave) communications systems. However, channel estimation is very challenging because limited measurements can be observed in the baseband.
\par
Several channel estimation approaches have been proposed \cite{Gao2017,Yang2017, Gao2019}. The support detection approach in \cite{Gao2017} utilizes the structural characteristics of the mmWave beamspace channel to reduce the pilot overhead while maintaining reliable performance. In \cite{Yang2017}, the channel matrix is regarded as a 2D image, which converts channel estimation into image reconstruction. These approaches have been developed for narrowband mmWave systems and therefore ignore the \emph{beam squint} effect in wideband mmWave systems. In \cite{Gao2019}, the successive support detection (SSD) technique based on applying successive interference cancelation can estimate the wideband channels. Its performance can be further improved if several essential characteristics of mmWave channels, such as sparsity and channel correlation between adjacent antennas and subcarriers, are considered.
\par
Recently, deep learning (DL) has been applied to physical layer communications \cite{Qin2019, He2019, Wen2018, Ye2018, He2018_1, He2018}, including channel state information feedback \cite{Wen2018}, signal detection\cite{Ye2018, He2018_1}, and channel estimation \cite{He2018}. As a prevailing approach, model-driven DL combines expert knowledge in wireless communications with DL and has been proved to be very effective \cite{He2019, He2018}. For channel estimation, a model-driven DL network unfolds the approximate message passing algorithm and can estimate narrowband beamspace mmWave channels \cite{He2018}. A learned denoising-based generalized expectation consistent (LDGEC) signal recovery network is developed for wideband beamspace channel estimation \cite{He2020}. These DL-based approaches can obtain better performance than traditional compressed sensing (CS)-based algorithms because of the powerful ability of DL. However, they only regard the mmWave channel as an image and use a complex denoising convolutional neural network to learn the channel structure. The inherent sparsity property of mmWave channel is not utilized, resulting in many learnable parameters and high complexity.
\par
In this article, we investigate a model-driven DL-based channel estimator for the lens-based wideband beamspace MIMO-orthogonal frequency-division multiplexing (OFDM) system. The network structure is obtained by unfolding the iterative shrinkage thresholding algorithm (ISTA) with a few learnable parameters. Different from other DL-based channel estimator using complex image denoiser, the proposed channel estimator learns to transform the beam-frequency domain mmWave channel into the domain with sparse features, thereby enables us to adopt a simple denoiser with few trainable parameters. Leveraging a small number of key parameters in the network, we further introduce another neural network named \emph{HyperNet} to generate learnable parameters online automatically. Simulation results show that the proposed approach can significantly outperform the state-of-the-art DL-based algorithms with lower complexity and fewer parameters and present fast adaptation to the channel environment.

\section{System Model and Problem Formulation}
\label{sec_system_model}
\par
In this section, we first present the lens-based wideband beamspace MIMO-OFDM systems. After introducing the beam squint effect, we formulate the beamspace channel estimation as a compressed image recovery problem.
\subsection{mmWave Beamspace Channel Model}
\begin{figure*}[htbp!]
  \centering
  \includegraphics[width=0.62\textwidth]{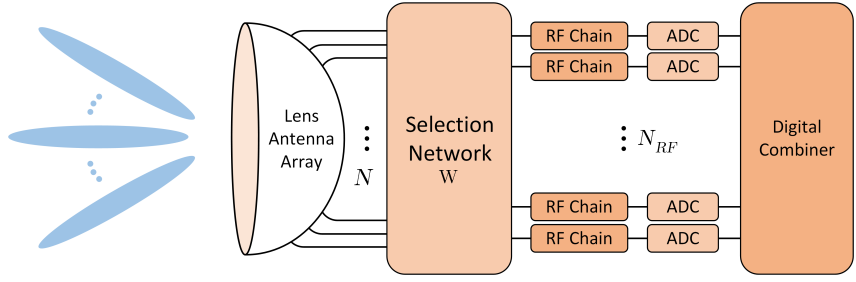}
  \caption{.~~Lens-based wideband beamspace MIMO-OFDM system.}
  \vspace{-0.2em}
  \label{fig_system_model}
\end{figure*}
 As illustrated in Fig.\,\ref{fig_system_model}, we consider a lens-based wideband beamspace MIMO-OFDM system. The base station (BS) equipped with an $N$-element lens antenna array and $N_{RF}$ RF chains simultaneously serves a single-antenna user. Considering the widely used Saleh-Valenzuela channel model\cite{Gao2019}, the spatial channel $\mathbf{h}_m \in \mathbb{C} ^{N\times 1}$ at sub-carrier $m$ is given by
\begin{equation}
  \mathbf{h}_{m}=\sqrt{\frac{1}{L}} \sum_{l=1}^{L} \alpha_{l} e^{-j 2 \pi \tau_{l} f_{m}} \mathbf{a}\left(\phi_{l, m}\right),
  \label{equ_h}
\end{equation}
for $m=1,2,\dots, M$ where $L$ is the number of resolvable paths, $\alpha_{l}$ and $\tau_{l}$ are the complex gain and the time delay of the $l$-th path, respectively. Furthermore, $\mathbf{a}(\phi_{l,m})$ is the array response vector and $\phi_{l,m}$ is the spatial direction at sub-carrier $m$ defined as
\begin{equation}\label{equ_phi}
  \phi_{l,m} = \frac{f_{m}}{c}d \sin\theta_{l},
\end{equation}
where $f_{m} = f_{c} + \frac{f_{b}}{M}(m-1-\frac{M-1}{2})$ is the frequency of sub-carrier $m$ with $f_{c}$ and $f_{b}$ representing the carrier frequency and bandwidth, respectively, $c$ is the speed of light, $\theta_{l}$ is the physical direction, and $d$ is the antenna spacing, which is usually designed according to the carrier frequency as $ d =0.5\cdot c/f_{c}$. A large $f_{b}$ in wideband system causes $\phi_{l,m}$ to vary greatly in different frequency of subcarrier, and results in different spatial channel $\mathbf{h}_m$, which is called beam squint effect. When the BS is equipped with a uniform linear lens antenna array, the array response vector $\mathbf{a}(\phi_{l,m})$ can be represented as
\begin{align}\label{equ_ULA}
  \mathbf{a}(\phi_{l,m}) & = e^{-j2\pi\phi_{l,m}\mathbf{p}_a},
\end{align}
where $\mathbf{p}_{a}=\left[-\frac{N-1}{2},-\frac{N+1}{2}, \ldots, \frac{N-1}{2}\right]^\top$ represents the index of different antennas. Lens antenna array as shown in Fig.\,\ref{fig_system_model} can transform the spatial domain in (\ref{equ_h}) into the beamspace domain. Specifically, lens antenna array plays the role of an $N\times N$-element spatial discrete Fourier transform (DFT) matrix $\mathbf{F}$. Accordingly, the wideband beamspace channel $\tilde{\mathbf{h}}_{m}$ at sub-carrier $m$ can be expressed as
\begin{equation}
  \tilde{\mathbf{h}}_{m} = \mathbf{F}^{H} \mathbf{h}_{m} = \sqrt{\frac{1}{L} } \sum_{l=1}^{L}\alpha_{l}e^{-j2\pi\tau_{l}f_{m}}\tilde{\mathbf{c}}_{l,m},
  \label{equ_beam}
  \vspace{-0.1em}
\end{equation}
where $\tilde{\mathbf{c}}_{l,m}$ denotes the $l$-th path component at sub-carrier $m$ in the beamspace domain and determined by $\phi_{l,m}$ as
\begin{align}\label{equ_c}
  \tilde{\mathbf{c}}_{l,m} & = \mathbf{F}^{H}\mathbf{a}(\phi_{l,m}) \\  \nonumber
    & = [\Xi(\phi_{l,m}-\bar{\phi}_{1}), \Xi(\phi_{l,m}-\bar{\phi}_{2}),\ldots,\Xi(\phi_{l,m}-\bar{\phi}_{N})]^{\top},
\end{align}
\vspace{-0.1em}
where $\Xi(x) = \frac{\sin N \pi x}{\sin \pi x}$ is the Dirichlet sinc function and $\bar{\phi}_{n}=\frac{1}{N}(n-\frac{N+1}{2})$ for $n=1,2,\ldots,N$ are the spatial directions pre-defined by the lens antenna array.
\vspace{-0.1em}
\subsection{Problem Formulation}
\par
The received signal vector $\mathbf{y}_{m,q}\in \mathbb{C}^{N_{RF}\times 1}$ at the BS can be written as
\begin{equation}
  \mathbf{y}_{m, q}=\mathbf{W}_{q} \tilde{\mathbf{h}}_{m} s_{m, q}+\mathbf{W}_{q} \mathbf{n}_{m, q},
\end{equation}
where $s_{m,q}$ is the pilot transmitted at sub-carrier $m$ in instant $q$ for $m = 1,2,\dots,M $ and $q = 1,2,\dots,Q$, and $\mathbf{n}_{m, q} \sim \mathcal{N}_{\mathbb{C}}\left(\mathbf{0}, \sigma^{2} \mathbf{I}\right)$ represents a Gaussian noise vector. $\mathbf{W}_{q} \in \mathbb{C}^{N_{RF}\times N}$ is the adaptive selection network that is fixed for different sub-carriers. The pilot signal is known at the receiver side and we set $s_{m,q}=1$ for convenience. The received signal $\bar{\mathbf{y}}_{m}$ in $Q$ instants is given by
\begin{equation}\label{equ_y_all}
  \bar{\mathbf{y}}_{m}=[\mathbf{y}_{m,1}^{\top}, \ldots, \mathbf{y}_{m,Q}^{\top}]^{\top}=\bar{\mathbf{W}}\tilde{\mathbf{h}}_{m}+\mathbf{n}_{m}^{\text{eq}},
\end{equation}
where $\bar{\mathbf{W}} = [\mathbf{W}_{1}^{\top}, \mathbf{W}_{2}^{\top},\ldots,\mathbf{W}_{Q}^{\top}]^{\top} \in \bbC^{QN_{RF}\times N} $ and $\mathbf{n}_{m}^{\text{eq}}=[(\mathbf{W}_{1}\mathbf{n}_{m,1})^{\top}, \ldots, (\mathbf{W}_{Q}\mathbf{n}_{m,Q})^{\top}]^{\top}$. In this article, low-cost $one$-bit phase shifters are utilized in the adaptive selection network $\mathbf{W}_{q}$. Therefore, the elements of $\bar{\mathbf{W}}$ are randomly selected from the set $\frac{1}{\sqrt{Q N_{RF}}}\{-1,+1\}$ with equal probability.
\par
Because the beamspace channel vectors at different subcarriers are correlated through lens antenna array response vector $\mathbf{a} \left(\phi_{l, m} \right)$, which is highly similar to a $2$D natural image. By stacking $M$ beamspace channel vectors into a matrix and transforming the complex values into real values, we have the following signal recovery model,
\begin{equation}\label{equ_linearmodel}
  \mathbf{Y} = \bar{\mathbf{W}}\mathbf{H}+\mathbf{N},
\end{equation}
where $\mathbf{Y} = [\text{Re}(\bar{\mathbf{y}}_{1},\bar{\mathbf{y}}_{2},\ldots,\bar{\mathbf{y}}_{M}), \text{Im}(\bar{\mathbf{y}}_{1},\bar{\mathbf{y}}_{2},\ldots,\bar{\mathbf{y}}_{M})]\in \mathbb{R}^{QN_{RF}\times 2M}$ is the received signal in the real value domain, $\mathbf{H} = [\text{Re}(\tilde{\mathbf{h}}_{1},\tilde{\mathbf{h}}_{2},\ldots,\tilde{\mathbf{h}}_{M}), \text{Im}(\tilde{\mathbf{h}}_{1},\tilde{\mathbf{h}}_{2},\ldots,\tilde{\mathbf{h}}_{M})] \in \mathbb{R}^{N\times 2M}$, and $\mathbf{N} = [\text{Re}(\mathbf{n}_{1}^{\mathrm{eq}}, \mathbf{n}_{2}^{\mathrm{eq}},\ldots,\mathbf{n}_{M}^{\mathrm{eq}}), \text{Im}(\mathbf{n}_{1}^{\mathrm{eq}}, \mathbf{n}_{2}^{\mathrm{eq}},\ldots,\mathbf{n}_{M}^{\mathrm{eq}})]$. If we regard the beam-frequency matrix $\mathbf{H}$ as a $2$D natural image, many compressed image recovery methods can be used here for beamspace channel estimation, which enables us to develop a model-driven DL-based channel estimation network.
\section{Proposed Channel Estimation Network}
\label{sec_LISTA_CE}
In this section, we propose the learnable ISTA-based channel estimator (LISTA-CE) for beamspace channel estimation. After introducing the traditional ISTA algorithm, we present the network structure.
\subsection{Traditional ISTA Algorithm}
\par
ISTA is a classical signal recovery algorithm in CS \cite{Warwick1864}. As we regard the beam-frequency channel as a $2$D picture, the beamspace channel estimation problem can be regarded as the compressed image recovery problem and the ISTA algorithm can be applied to estimate the beamspace channel.
\par
ISTA algorithm recovers the signal by updating the following steps iteratively,
\begin{equation}
  \mathbf{R}^{(t)}=\mathbf{\hat{H}}^{(t - 1)}-\rho\bar{\mathbf{W}}^{\top}\left(\bar{\mathbf{W}} \mathbf{\hat{H}}^{(t-1)}-\mathbf{Y}\right)
  \label{equ_ISTA_r}
\end{equation}
\begin{equation}
  \mathbf{\hat{H}}^{(t)}=\underset{\mathbf{H}}{\arg \min } \frac{1}{2}\left\|\mathbf{R}^{(t)} - \mathbf{H}\right\|_{2}^{2}+\lambda\|\mathbf{\Psi} \mathbf{H}\|_{1}
  \label{equ_ISTA_x}
\end{equation}
where $\boldsymbol{\Psi}$ is the transformation matrix that makes the $\mathbf{\Psi} \mathbf{H}$ sparse in some transform domain, $t$ is the iteration index, and $\rho$ is the iteration stepsize. The two update steps (\ref{equ_ISTA_r}) and (\ref{equ_ISTA_x}) are mainly used to perform gradient descent and optimization process, respectively. Therefore, $\mathbf{R}^{(t)}$ can be regarded as the noisy signal and $(\ref{equ_ISTA_x})$ is used to perform denoising on $\mathbf{R}^{(t)}$ regularized by $\mathbf{\Psi} \mathbf{H}$.
\subsection{LISTA-CE-based Channel Estimator}
\begin{figure}[t]
  \begin{minipage}{3in}
    \centerline{\includegraphics[width=2.85in]{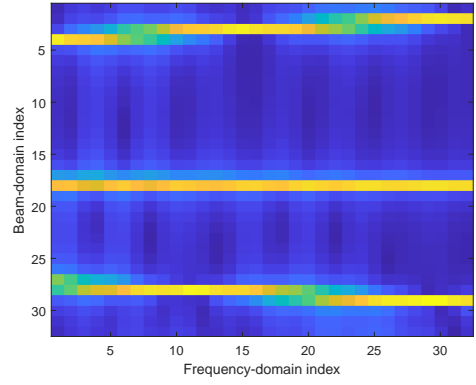}}
    \centerline{(a) Beam-frequency domain} \label{fig_channel_a}
  \end{minipage}
  \hfill
  \begin{minipage}{3in}
    \centerline{\includegraphics[width=2.85in]{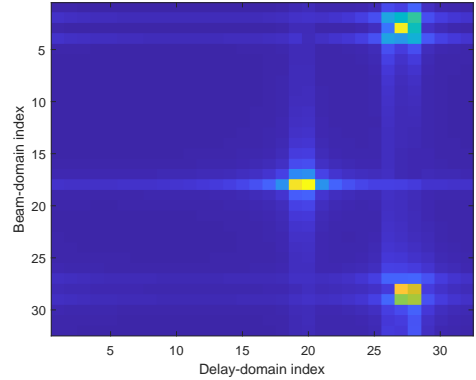}}
    \centerline{(b) Beam-delay domain} \label{fig_channel_b}
  \end{minipage}
  \caption{.~~Pseudo-color of the mmWave channel with $L = 3$, $M = 32$, $N = 32$, $f_c = 28\text{ GHz}$ and $f_b = 4\text{ GHz}$. }
  \vspace{-1em}
  \label{fig_channel}
\end{figure}
\par
Although ISTA is an effective algorithm to solve the compressed signal recovery problem, many numerical parameters, such as $\lambda$ and $\rho$, need to be selected manually. Furthermore, choosing a proper transformation matrix $\mathbf{\Psi}$ is challenging. Usually, the DFT matrix is utilized but it is not an accurate sparse transformation matrix for mmWave channel, especially for a widedband system with the beam squint effect. For example, the pseudo-color map of the beam-delay domain channel in Figure.\,\ref{fig_channel}(b) is obtained by taking DFT for the beam-frequency matrix in Fig.\,\ref{fig_channel}(a). From the figure, the beam-delay domain is not exactly sparse in the transform domain, which motivates us to utilize the DL to learn the sparse transformation matrix $\mathbf{\Psi}$.
\begin{figure*}[htbp!]
  \centering
  \includegraphics[width=0.9\textwidth]{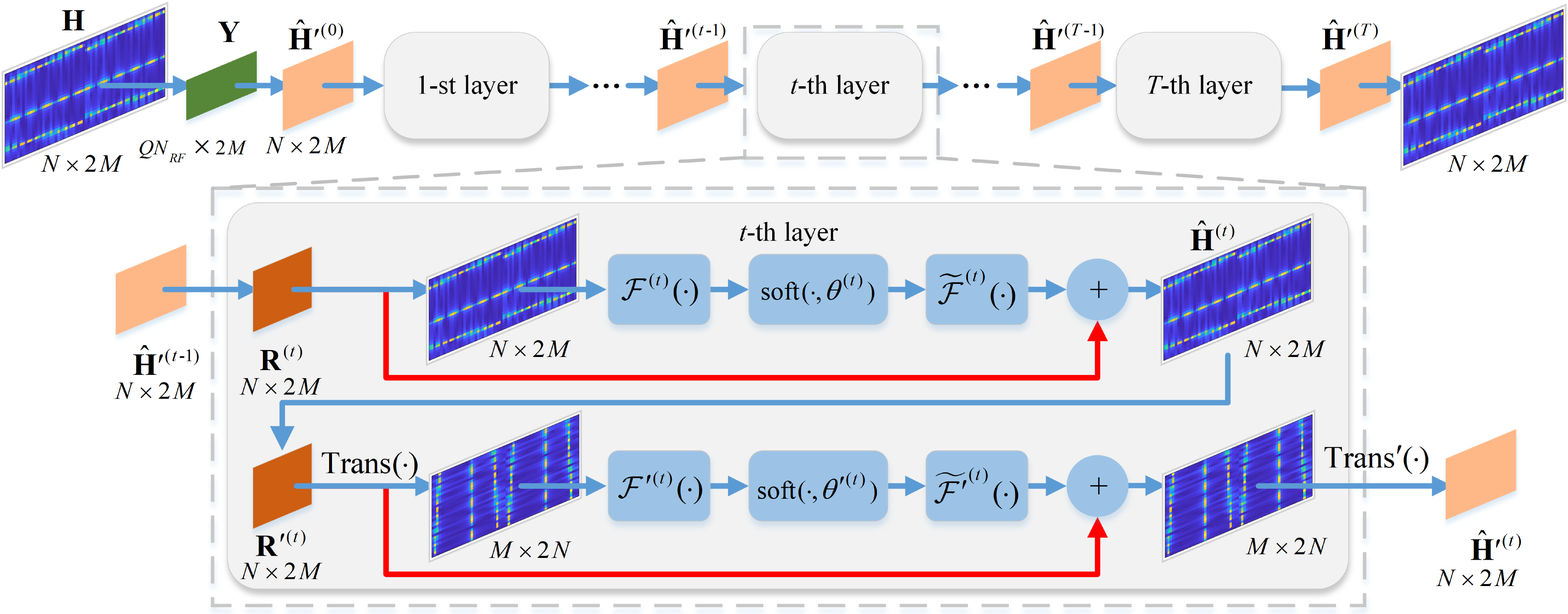}
  \caption{.~~Illustration of the proposed LISTA-CE framework.}
  \vspace{-0.5em}
  \label{fig_LISTA_structure}
\end{figure*}
\par
To solve the problem, we develop the LISTA-CE network. The structure of LISTA-CE is illustrated in Fig.\,\ref{fig_LISTA_structure}, which is a revised version of ISTA algorithm by adding learnable
variables. The network consists of $T$ cascade layers and each has the same structure with four modules: Module $\mathbf{R}^{(t)}$, Module $\mathbf{\hat{H}}^{(t)}$, Module $\mathbf{R}'^{(t)}$ and Module $\mathbf{\hat{H}}'^{(t)}$. The input of the LISTA-CE is the received signal $\mathbf{Y}$, $\mathbf{\hat{H}}^{\prime(0)} = 0$ and the final output $\mathbf{\hat{H}}^{\prime(T)}$ is the estimation of $\mathbf{H}$. 
\par
In LISTA-CE, Module $\mathbf{R}^{(t)}$ and Module $\mathbf{\hat{H}}^{(t)}$ perform channel transformation and denoising in the frequency domain while Module $\mathbf{R}'^{(t)}$ and Module $\mathbf{\hat{H}}'^{(t)}$ perform channel transformation and denoising in the beam domain, respectively. Following are details on the four modules.
\begin{itemize}
\item[$\bullet$] \textbf{Module $\mathbf{R}^{(t)}$:} $\mathbf{R}^{(t)}$ can be written as
\begin{equation}
  \mathbf{R}^{(t)}=\mathbf{\hat{H}}^{(t - 1)}-\rho^{(t)} \bar{\mathbf{W}}^{\top}\left(\bar{\mathbf{W}} \mathbf{\hat{H}}^{\prime(t-1)}-\mathbf{Y}\right),
  \label{equ_iter_r1}
\end{equation}
which is obtained by deep unfolding the ISTA algorithm into network, and $\mathbf{\hat{H}}^{\prime(t-1)}$ is caculated in Module $\mathbf{\hat{H}}'^{(t - 1)}$, which will be introduced later. Different from (\ref{equ_ISTA_r}) in the ISTA algorithm, the stepsize $\rho^{(t)}$ in ($\ref{equ_iter_r1}$) is learned from data and not shared for each layer. If $\rho^{(t)}$ is fixed and shared for all layers (\ref{equ_iter_r1}) is reduced to (\ref{equ_ISTA_r}).

\item[$\bullet$] \textbf{Module $\mathbf{H}^{(t)}$:} Denote $\mathcal{F}^{(t)}(\mathbf{R}^{(t)}) = \mathbf{B}^{(t)}\cdot\text{ReLU}\left(\mathbf{A}^{(t)}\cdot \left(\mathbf{R}^{(t)}\right)^{\top}\right)$ as the sparse transformation for $\mathbf{R}^{(t)}$, where $\mathbf{A}^{(t)} \in \mathbb{R}^{w_1\times 2M} $ and $\mathbf{B}^{(t)} \in \mathbb{R}^{w_2\times w_1}$ are two learnable matrices. As proven in \cite{Zhang2018}, $||\mathcal{F}^{(t)}(\mathbf{R}^{(t)}) - \mathcal{F}^{(t)}(\mathbf{H})||_2 ^2 \approx \alpha^{(t)} ||\mathbf{R}^{(t)} - \mathbf{H}||_2 ^2$. Therefore, we can rewrite (\ref{equ_ISTA_x}) as
\begin{equation}
  \begin{aligned}
    \hat{\mathbf{H}}^{(t)}=& \underset{\mathbf{H}}{\arg \min }\left(\frac{1}{2}\left\|\mathcal{F}^{(t)}\left(\mathbf{R}^{(t)}\right)-\mathcal{F}^{(t)}(\mathbf{H})\right\|_{2}^{2}\right.\\
    &\left.+\beta^{(t)}\left\|\mathcal{F}^{(t)}(\mathbf{H})\right\|_{1}\right)
    \end{aligned}
  \label{equ_iter_x1_trans}
\end{equation}
where $\beta^{(t)} = \lambda^{(t)} \times \alpha^{(t)}$. The optimization problem in (\ref{equ_iter_x1_trans}) can be interpreted as the denoising problem and solved by the soft denoiser, $\text{soft}(x,\theta) = \mathrm{sign(}x)(\mathrm{max}(0,|x| - \theta)$. Furthermore, $\widetilde{\mathcal{F}}^{(t)}(\cdot)$ is denoted as the inverse transformation for converting the signal from the sparse transform domain to the beam-frequency domain. The network structure of $\widetilde{\mathcal{F}}^{(t)}(\cdot)$ is symmetric to that of $\mathcal{F}^{(t)}(\cdot)$, which is composed of two fully connected layers. Therefore, the closed-form solution for (\ref{equ_iter_x1_trans}) is given by
\begin{equation}
  \mathbf{\hat{H}}^{(t)}=\widetilde{\mathcal{F}}^{(t)}\left(\text{soft}\left(\mathcal{F}^{(t)}\left(\mathbf{R}^{(t)}\right), \theta^{(t)}\right)\right),
  \label{equ_iter_x1}
\end{equation}
where $\theta^{(t)}$ represents the threshold of the soft denoiser in the $t$-th layer.
\par
To further speed up the convergence speed, we introduce the concept of residual learning \cite{He2016} into LISTA-CE. As shown in Fig.\,\ref{fig_LISTA_structure}, the red lines represent direct connection of residual learning. Therefore, we can rewrite (\ref{equ_iter_x1}) as
\begin{equation}
  \mathbf{\hat{H}}^{(t)} = \mathbf{R}^{(t)} + \widetilde{\mathcal{F}}^{(t)}\left(\text{soft}\left(\mathcal{F}^{(t)}\left(\mathbf{R}^{(t)}\right), \theta^{(t)}\right)\right).
  \label{equ_iter_x1_residual}
\end{equation}

\item[$\bullet$] \textbf{Module $\mathbf{R}'^{(t)}$:} Module $\mathbf{R}'^{(t)}$ is similar to Module $\mathbf{R}^{(t)}$ but with $\mathbf{\hat{H}}^{(t)}$ as input and $\rho'^{(t)}$ as stepsize, that is, 
\begin{equation}
  \mathbf{R}^{\prime(t)}=\mathbf{\hat{H}}^{(t)}-\rho'^{(t)} \bar{\mathbf{W}}^{\top}\left(\bar{\mathbf{W}} \mathbf{\hat{H}}^{(t)}-\mathbf{Y}\right).
  \label{equ_iter_r2}
\end{equation}

\item[$\bullet$] \textbf{Module $\mathbf{\hat{H}}'^{(t)}$:} 
To perform channel transformation and denoising in the beam domain, we need to transpose $\mathbf{R}^{\prime(t)}$ firstly because the left multiplication operation in the sparse transformation represents the transformation of the columns of the matrix. Denote the transpose operations as $\text{Trans}(\mathbf{R}^{\prime(t)}) = [\mathbf{R}^{\prime(t)}(:, 1:M)^\top, \mathbf{R}^{\prime(t)}(:, M+1:2M)^\top]$ and the inverse operation $\text{Trans}'(\cdot)$. Therefore, $\mathbf{\hat{H}}^{\prime(t)}$ can be denoted as
\begin{equation}
  \begin{aligned}
  \mathbf{\hat{H}}^{\prime(t)} &= \mathbf{R}'^{(t)} + \\&\text{Trans}'\left(\widetilde{\mathcal{F}}'^{(t)}\left(\text{soft}\left(\mathcal{F}'^{(t)}\left(\text{Trans}\left(\mathbf{R}'^{(t)}\right)\right), \theta'^{(t)}\right)\right)\right).
  \end{aligned}
  \label{equ_iter_x2}
\end{equation}
The structures of $\mathcal{F}'^{(t)}(\cdot)$ and $\widetilde{\mathcal{F}}'^{(t)}(\cdot)$ are similar to $\mathcal{F}^{(t)}(\cdot)$ and $\widetilde{\mathcal{F}}^{(t)}(\cdot)$, respectively, but with different weights. 
\end{itemize}
\par
From the above discussion, the learnable variables in\linebreak[2]LISTA-CE network are $\Theta = \{\rho^{(t)}, \rho'^{(t)}, \theta^{(t)}, \theta'^{(t)}, \mathcal{F}^{(t)}(\cdot), $ $\widetilde{\mathcal{F}}^{(t)}(\cdot), \mathcal{F}'^{(t)}(\cdot), \widetilde{\mathcal{F}}'^{(t)}(\cdot)\}_{t = 1} ^{T}$.

\section{Adaptive LISTA-CE Network}
For DL-based channel estimator, how to adapt to channel environment rapidly is significantly important. In this section, we design the \emph{LISTA-CEHyper} network that introduces the \emph{HyperNet} network into LISTA-CE network to adjust several trainable variables to adapt the channel environment. After introducing the network structure, we analyze the complexity of different algorithms.
\vspace{-0.1em}
\subsection{LISTA-CEHyper Network}\label{sec_LISTA_Hyper}
\begin{figure*}[htbp!]
  \centering
  \includegraphics[width=0.85\textwidth]{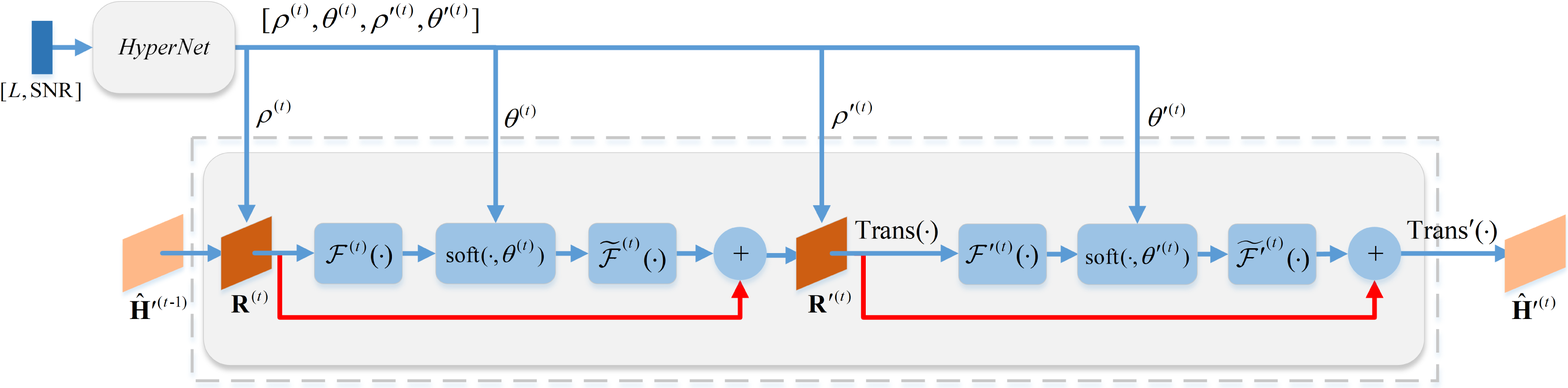}
  \caption{.~~Illustration of the $t$-th layer of our proposed LISTA-CEHyper framework.}
  \label{fig_LISTA_CEHyper_structure}  
  \vspace{-0.5em}  
\end{figure*}
Although the LISTA-CE network has superior channel estimation performance, the training and test environment mismatch will result in performance degradation. This is because the trainable variables obtained in specific settings are not suitable for another. For example, the LISTA-CE trained with the data generated from $L = 3$ and SNR$ = 10$ dB will have performance degradation when tested at $L = 2$ and SNR$ = 15$ dB. To solve the problem, we introduce the hypernetwork \cite{Johnson2014} into the LISTA-CE. In \cite{Johnson2014}, the hypernetwork can generate a small number of weights for another network based on important features. As $\{\rho^{(t)},\rho'^{(t)}, \theta^{(t)}, \theta'^{(t)}\}_{t=1}^{T}$ are significantly related to the number of resolvable paths $L$ and SNR value, $\mathbf{s} = \{L, \text{SNR}\} \in \mathbb{R}^{2\times 1}$ and $\{\rho^{(t)}, \rho'^{(t)}, \theta^{(t)}, \theta'^{(t)}\} _{t=1}^{T} \in \mathbb{R}^{4T\times 1}$ are chosen as the input and output for \emph{HyperNet}, respectively\footnote{Noth that the number of resolvable paths $L$ can be obtained by simple channel estimator in online stage.}. Denote $\bm{\rho} = \{\rho^{(t)}, \rho'^{(t)}\}_{t = 1} ^{T}$ and $\bm{\theta} = \{\theta^{(t)}, \theta'^{(t)}\}_{t = 1} ^{T}$ for convenience. Three fully connected layers are used for \emph{HyperNet}, which can be written as
\begin{equation}
  g(\mathbf{s}) = \mathbf{W}_3 \cdot \text{ReLU}(\mathbf{W}_2 \cdot \text{ReLU}(\mathbf{W}_1\cdot \mathbf{s})),
\end{equation}
where $\mathbf{W}_1 \in \mathbb{R}^{d\times 2}, \mathbf{W}_2 \in \mathbb{R}^{d\times d}, \text{and }\mathbf{W}_3 \in \mathbb{R}^{4T\times d}$.
\par
We first train LISTA-CE with all possible $L$ and SNR and obtain an average model called \emph{LISTA-CEAver}. Then, the learnable sparse transformations are fixed and \emph{HyperNet} is introduced to generate $\{\bm{\rho}, \bm{\theta}\}$ in the online deployment phase. We denote LISTA-CE combined with \emph{HyperNet} as \emph{LISTA-CEHyper} as shown in Fig.\,\ref{fig_LISTA_CEHyper_structure}.
\vspace{-0.2em}
\begin{table}[htbp]
  \caption{Complexity Analysis}
  \label{tab_complexity}
    \centering
    \begin{tabular}{ccc}
     \toprule
                & Parameters  & \makecell[c]{Computational \\Complexity}\\
     \midrule
     LISTA-CE & $1.65\times 10^5$      & $\mathcal{O}(QN_{RF}NM)$ \\
     LISTA-CEHyper & $1.67\times 10^5$  & $\mathcal{O}(QN_{RF}NM)$ \\
     LDGEC & $5.19\times 10^5$          & $\mathcal{O}(MN^3)$ \\
     ISTA & 2                           & $\mathcal{O}(QN_{RF}NM)$\\
     ISTA-Net$^+$ & $3.76\times10^4$       & $\mathcal{O}(MNk^2C_{\text{in}}C_{\text{out}})$\\
     SSD & 0                            & $\mathcal{O}(MN_{RF}QL^2\Omega ^2)$ \\
     OMP & 0                            & $\mathcal{O}(MN_{RF}QL^3\Omega ^3)$\\
  \bottomrule
  \end{tabular}
\end{table}
\vspace{-0.5em}
\subsection{Complexity Analysis}\label{sec_result}
We compare the complexity of LISTA-CE with other channel estimators, including LDGEC\cite{He2020}, ISTA\cite{Warwick1864}, ISTA-Net$^+$\cite{Zhang2018}, SSD\cite{Gao2019} and orthogonal matching pursuit (OMP)\cite{Venugopal2017}. As shown in Table \ref{tab_complexity}, the complexities of the SSD and the OMP algorithms are $O(MN_{RF}QL^2\Omega ^2)$ and $O(MN_{RF}QL^3\Omega ^3)$, respectively, where $\Omega$ is the beam windows and much smaller than $N$ ($\Omega$ is assumed as $\Omega=4$ when $N = 256$). Compared to the DL-based LDGEC network and ISTA-Net$^+$, the proposed LISTA-CE and LISTA-CEHyper have lower computational complexity because the complexity of LISTA-CE is mainly determined by matrix multiplication, which is $O(QN_{RF}NM)$, while the complexity of LDGEC network is $O(MN^3)$ determined by matrix inversion. The computational complexity of ISTA-Net$^+$ is $\mathcal{O}(MNk^2C_{\text{in}}C_{\text{out}})$, dominated by convolution operation, where $k$ is the size of the filter and $C_{\text{in}}$ and $C_{\text{out}}$ are the numbers of input and output channels of convolution, respectively, and $C_{\text{in}}=C_{\text{out}}=32 $ in \cite{Zhang2018}. Although LISTA-CE has more trainable parameters than ISTA-Net$^+$, it can achieve better performance shown by simulation results in the next section.
\section{Simulation Results}
\begin{figure}[htbp!]
  \vspace{-0.2em}
  \centering
  \includegraphics[width=2.85in]{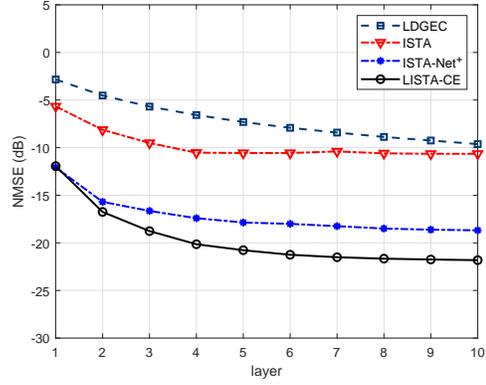}
  \caption{.~~Convergence analysis of the LISTA-CE and other channel estimation algorithms.}
  \vspace{-0.6em}
  \label{fig_NMSE_convergence}
\end{figure}
\vspace{-0.2em}  
\begin{figure}[htbp!]
  \centering
  \includegraphics[width=2.85in]{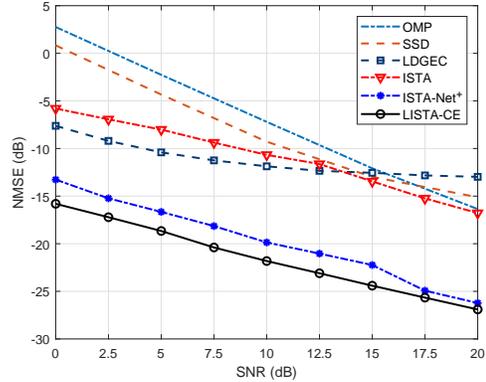}
  \caption{.~~NMSE analysis of the LISTA-CE and other channel estimation algorithms.}
  \label{fig_NMSE_SNR_Algorithm}
  \vspace{-0.8em}  
\end{figure}
In this section, we provide numerical results to show the performance of the proposed adaptive model-driven DL network for wideband beamspace channel estimation. The channel matrix is generated according to the system model (\ref{equ_h}) in Section \ref{sec_system_model}. We assume $N = 32$ and $N_{RF} = 8$. The carrier frequency is $f_c = 28$ GHz, the bandwidth is $f_b = 4$ GHz, and the number of subcarriers is $M = 32$. The channel matrix is generated with $L = 3$, $\theta_{l} \sim \mathcal{U}(-\pi/2, \pi/2)$, $ \tau_{l} \sim \mathcal{U}(0, 20\text{ ns})$ and the maximum delay $\tau_{\max} = 20$ ns.
\begin{figure}[htbp!]
  \centering
  \includegraphics[width=2.85in]{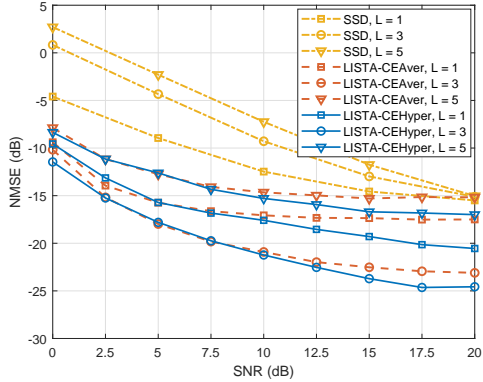}
  \caption{.~~NMSE analysis of the LISTA-CEHyper and LISTA-CEAver in different SNRs.}
  \label{fig_NMSE_SNR_Hyper}
  \vspace{-0.6em}
\end{figure}
\par
The training, validation, and testing datasets contain 10000, 1280, and 2560 samples, respectively. We set the dimensions of matrices, $w_1$, $w_2$, and $d$ in LISTA-CE and LISTA-CEHyper to 128, 256, and 128, respectively, and set the batch size to 64. We generate the adaptive selection network, $\bar{\mathbf{W}}$, for each batch. The LISTA-CE is trained using the Adam optimizer with learning rate $0.0001$. Furthermore, the learning rate is $0.00001$ when training the \emph{HyperNet} of LISTA-CEHyper. We use the normalized mean-squared error (NMSE) as the performance metric, which is defined as
\begin{equation}
  \mathrm{NMSE}=\mathbb{E}\left\{\|\mathbf{\hat{H}}^{\prime(T)}-\mathbf{H}\|_{2}^2/\|\mathbf{H}\|_{2}^2\right \},
  \label{equ_NMSE}
\end{equation}
\par
We firstly analyze the convergence of LISTA-CE with other channel estimation algorithms. As shown in Fig.\,\ref{fig_NMSE_convergence}, all the algorithms can converge within 10 layers except the LDGEC network. Specifically, the proposed LISTA-CE can converge within 7 layers. Therefore, the LISTA-CE network can be deployed with a small number of layers in the online stage.
\par
Fig.\,\ref{fig_NMSE_SNR_Algorithm} compares the NMSE performance of the LISTA-CE with other channel estimation algorithms. The LISTA-CE can outperform other CS-based algorithms, including SSD, ISTA, and OMP. Compared to the LDGEC network, the LISTA-CE has better NMSE performance and lower complexity shown in section \ref{sec_result} because LISTA-CE takes the sparsity of beamspace channel in the transform domain into consideration. Compared with ISTA-Net$^+$, the LISTA-CE has approximately $2.5$ dB performance gain and less computational complexity shown in section \ref{sec_result}. Furthermore, we utilize the DFT to transform the frequency domain into the delay domain and consider the ISTA-based channel estimator. The LISTA-CE can outperform the ISTA algorithm, as the beamspace channel in the learned sparse transform domain is much more sparse than in the delay domain.
\par
Finally, we investigate the NMSE performance of the SSD algorithm, LISTA-CEHyper and LISTA-CEAver in Fig.\,\ref{fig_NMSE_SNR_Hyper}. The LISTA-CEAver and LISTA-CEHyper are trained in $L = \{2,3,4\}$ and $\mathrm{SNR}=[5, 15]$ dB. After training, we test the two networks for $L=\{1,3,5\}$ and $ \mathrm{SNR}=[0, 20]$ dB, which contains $L$ and SNR that has not been encountered in the training procedure. From the figure, LISTA-CEHyper can outperform LISTA-CEAver and the SSD algorithm in most scenarios, especially at high SNR, which indicates the LISTA-CEHyper can quickly adapt to new channel environments. 
\vspace{-0.1em}

\section{Conclusion}
\vspace{-0.1em}
\par
In this paper, we developed a model-driven unfolding network called LISTA-CE for wideband beamspace MIMO-OFDM channel estimation. The network can utilize the sparsity of the mmWave channel to reduce the complexity and the number of learnable variables. To adapt to the channel environment rapidly, the LISTA-CEHyper has been proposed by introducing the hypernetwork into LISTA-CE. Simulation results demonstrated that the LISTA-CE can outperform CS-based and existing DL-based channel estimation algorithms and the LISTA-CEHyper network can rapidly adapt to new environments.
\vspace{-0.2em}
\section*{Acknowledgment}
\vspace{-0.3em}
This work was supported in part by the National Natural Science Foundation of China under Grant 61941104. 
\vspace{-0.2em}

\bibliographystyle{ieeetran}
\bibliography{referencePaper}
\end{document}